\documentclass[conference]{IEEEtran}
\usepackage[marginal]{footmisc}

\usepackage{geometry}
\usepackage{tikz}
\usepackage{circledsteps}
\usepackage{booktabs}
\usepackage{multirow}
\usepackage{multicol}
\usepackage{circledsteps}
\usepackage{makecell}
\usepackage{subfigure}
\usepackage{hyperref}
\usepackage[normalem]{ulem}
\usepackage{cite}
\usepackage{amsmath,amssymb,amsfonts}
\usepackage{algorithm}
\usepackage{algorithmic}
\usepackage{graphicx}
\usepackage{textcomp}
\usepackage{xcolor}
\usepackage{amssymb}
\usepackage{soul}
\usepackage[normalem]{ulem}
\begin{document}

\title{\Large\bf AgenticTCAD: A LLM-based Multi-Agent Framework for Automated TCAD Code Generation and Device Optimization}

\author{
    \IEEEauthorblockN{Guangxi Fan$^{1,2}$, Tianliang Ma$^{1,2}$, Xuguang Sun$^{1,2}$, Xun Wang$^{1,2}$, Kain Lu Low$^{3}$*, and Leilai Shao$^{1,2}$*}
    \IEEEauthorblockA{$^{1}$State Key Laboratory of Micro-nano Engineering Science\\ $^{2}$Micro-nano Engineering Sciences Research Center\\School of Mechanical Engineering, Shanghai Jiao Tong University, Shanghai, China\\$^{3}$Xi'an Jiaotong–Liverpool University, Suzhou, China }
}

\maketitle



\makeatletter
\def\ps@IEEEtitlepagestyle{%
  \def\@oddfoot{\mycopyrightnotice}%
  \def\@evenfoot{}%
}
\makeatother
\def\mycopyrightnotice{%
  \begin{minipage}{\textwidth}
    \footnotesize
    ~ \hfill\\~\\
  \end{minipage}
  \gdef\mycopyrightnotice{}
}

\footnote{
\indent *Corresponding author:  Leilai Shao (leilaishao@sjtu.edu.cn); Kain Lu Low (kainlu.low@xjtlu.edu.cn). This work was supported by the National Key Research and Development Program of China: Design Technology Co-Optimization Methodology (2023YFB4402700), Shanghai Jiao Tong University AI for Engineering Initiative (26X010100040), Shanghai Municipal Science and Technology Major Project in collabration with Shanghai Artificial Intelligence Laboratory and Xi'an Jiaotong-Liverpool University Research Development Fund under Grant RDF-24-01-107. The authors would like to thank Prof. Libin Liu for the valuable discussions.}

{\small\bf Abstract--- With the continued scaling of advanced technology nodes, the design–technology co-optimization (DTCO) paradigm has become increasingly critical, rendering efficient device design and optimization essential. In the domain of TCAD simulation, however, the scarcity of open-source resources hinders language models from generating valid TCAD code. To overcome this limitation, we construct an open-source TCAD dataset curated by experts and fine-tune a domain-specific model for TCAD code generation. Building on this foundation, we propose \textbf{AgenticTCAD}, a natural language–driven multi-agent framework that enables end-to-end automated device design and optimization. Validation on a 2 nm nanosheet FET (NS-FET) design shows that \textbf{AgenticTCAD} achieves the International Roadmap for Devices and Systems (IRDS)-2024 device specifications within 4.2 hours, whereas human experts required 7.1 days with commercial tools.
}


\section{Introduction}
Over the past decades, the semiconductor industry has advanced by following Moore’s law, which has consistently increased transistor density, performance, and energy efficiency\cite{2022_75y}. As CMOS technology approaches its physical and economic limits, however, further scaling is hindered by challenges such as rising leakage currents, variability, and manufacturing costs, making it difficult to sustain historical improvements. To address these issues, the concept of design technology co-optimization (DTCO) has emerged, highlighting the need for joint optimization of process technology and circuit design to continue driving progress in integrated circuits and systems\cite{2024_dtco_ted}.

Within the DTCO paradigm, the technology-oriented dimension plays a central role, encompassing material exploration\cite{2023-hi-tcad-material}, device architecture innovation\cite{2022-ted-3nm-tcad}, and standard-cell layout optimization\cite{2024-iedm-cfet-layout}. Among these tasks, device design and optimization are especially critical, as transistor characteristics directly impact overall system performance. Consequently, Technology Computer-Aided Design (TCAD) simulation\cite{synopsys_tcad} is indispensable for guiding technology scaling and supporting effective DTCO.

TCAD simulation involves highly complex physical mechanisms, such like carrier transport, quantum confinement, and trap effects\cite{2022-iedm-mx2}. Manual device optimization under these conditions is extremely challenging, as the large design space coupled with intricate physics requires significant expertise and effort. In addition, TCAD tools require domain-specific coding languages to describe device structures, meshing, and simulation setups. Since semiconductor device research is often proprietary and lacks open-source references, writing such codes poses a significant barrier for researchers and further limits efficient adoption.

Large language models (LLMs) have recently demonstrated considerable potential in the design domain of DTCO\cite{2025-review-llm4eda}. They have been employed for code generation, enabling the automatic synthesis of design scripts and hardware descriptions directly from natural language inputs\cite{2025-iccad-origen-rtl}. Furthermore, LLMs have been investigated for automated circuit design, offering a pathway from high-level specifications to transistor-level design with reduced human involvement\cite{2024-dac-analogllm}.

The technology dimension of DTCO critically depends on device design and optimization. Sentaurus Structure Editor (SDE) and Sentaurus Device (SDevice) are widely used for device construction and electrical simulation, but their domain-specific and proprietary scripts leave the community without publicly accessible datasets for TCAD code generation. Current research efforts have so far focused mainly on SDE script generation. Early attempts explored nanowire device structure generation\cite{2024-nano-tcadllm}, but these did not include mesh construction, making direct electrical simulation infeasible. Subsequent work has investigated end-to-end code generation through fine-tuned LLM\cite{2025-chattcad}, yet this required translating SDE scripts into Python descriptions beforehand, which introduces additional complexity and has so far been limited to two-dimensional structures. Other approaches have employed agent-based methods to map layout descriptions into SDE scripts\cite{2025-lay2sde}, but these often depend on case-specific templates for in-context learning, reflecting the difficulty commercial LLMs still face in handling TCAD-specific syntax. 

These limitations highlight the absence of a unified framework that can transform natural-language specifications into both SDE and SDevice codes for end-to-end TCAD simulation, while also enabling iterative device parameter optimization with high physical interpretability. To address this gap, we propose \textbf{AgenticTCAD}, a domain-specific multi-agent LLM-based framework for automated device simulation and optimization, as illustrate in Fig. \ref{fig:overall_flow}. Our contributions are as follows:

\begin{figure*}[t]
\centering
\includegraphics[width=0.90\linewidth]{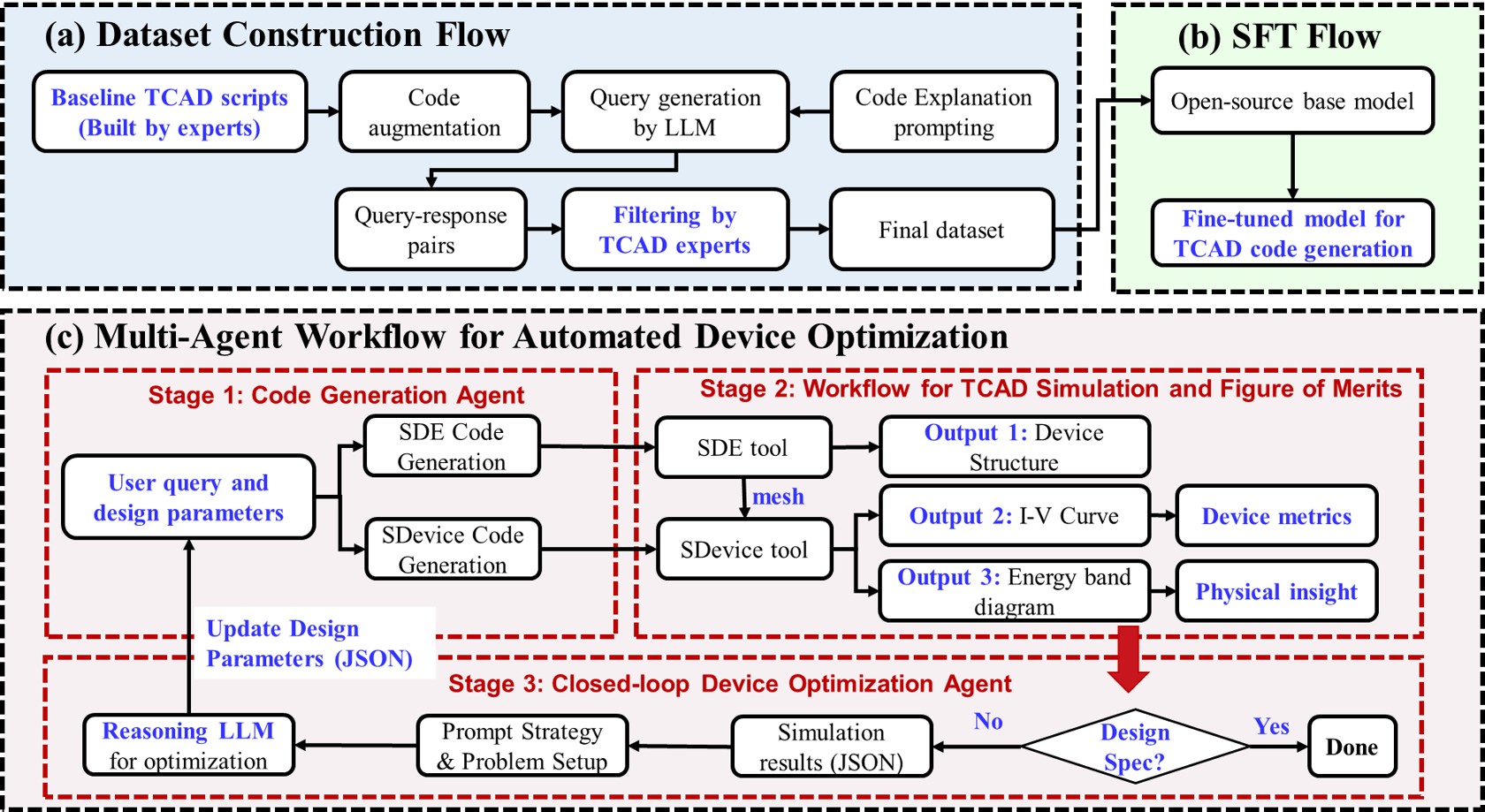}
\caption{Overall flow of TCADAgent. (a) Dataset construction flow; (b) Supervised fine-tuning (SFT) flow; (c) Multi-Agent flow for End-to-end device TCAD simulation and optimization based on natural language.}
\label{fig:overall_flow}
\end{figure*}

\begin{itemize}
\item We release the first open-source comprehensive TCAD dataset containing expert-crafted SDE and SDevice scripts, further expanded through an LLM-based augmentation pipeline, to support community research in this field. The dataset includes a wide spectrum of device structures in both two- and three-dimensional forms, various operating mechanisms including conventional MOSFETs as well as emerging junctionless and tunneling transistors, and multiple material systems including silicon and III–V semiconductors.
\item We fine-tune an open-source LLM to generate complete SDE and SDevice codes directly from natural language. The SDE codes specify geometry, doping profiles, contact definitions, and meshing strategies, while the SDevice codes define boundary conditions, physical models, and solver configurations, enabling direct end-to-end simulation.
\item We introduce \textbf{AgenticTCAD}, a multi-agent framework that integrates our fine-tuned TCAD code generation LLM with a commercial reasoning model, with each LLM agent using a tailored prompting strategy to realize automated device simulation and optimization. To demonstrate its effectiveness, we apply the framework to the optimization of design parameters for nanosheet field-effect transistors (NS-FETs) at advanced technology nodes. The framework also automatically extracts physical information from simulation results, such as energy band diagrams, to enhance the interpretability of the optimization flow.
\end{itemize}
\vspace{-0.08cm}

The remainder of this paper is organized as follows: Section II reviews related work on TCAD simulation, LLM-based code generation, and agent-based methods in EDA. Section III presents the proposed \textbf{AgenticTCAD} framework and its design details. Section IV reports experimental results and Section V concludes the paper. The code and dataset of \textbf{AgenticTCAD} are available at \href{https://github.com/guangxifan/llm4tcad_flow}{github}.

\section{preliminaries and related works}
\subsection{TCAD Simulation}
TCAD simulation has been widely applied in semiconductor device research and development to predict device behavior before fabrication. It supports a variety of material systems and device architectures, including advanced silicon-based technologies \cite{2022-ted-3nm-tcad}, III–V semiconductors\cite{2017-ted-tcad35}, and emerging devices such as carbon nanotube (CNT)\cite{2019-cnt-tcad}. In addition, TCAD enables comprehensive multiphysics analysis, capturing effects such as defects\cite{2022-iedm-mx2} and self-heating effects\cite{2024-ted-sh-tcad}. Owing to these capabilities, TCAD plays a crucial role in accelerating early-stage device design and optimization.
\subsection{EDA Code Generation with LLM}
Code generation is a key application of large language models (LLMs) and has been widely explored in mainstream programming language\cite{2024-cgen-tranditional}. Within DTCO’s design stage, LLMs have been applied to realize RTL code generation\cite{2025-iccad-origen-rtl, 2025-date-cgen-rl, 2025-date-rtl-cgen}, debugging \cite{2024-dac-rlt-fix, 2025-iccad-rtl-de} and optimization \cite{2025-iccad-rewrite}. In the back-end stage, researchers are investigating how LLMs can automatically construct physical design scripts, including flows such as placement and routing, clock-tree synthesis \cite{2024-tcad-chateda}. In the technology domain, there have been initial attempts to use LLMs for generating TCAD SDE input code \cite{2024-nano-tcadllm, 2025-chattcad, 2025-lay2sde} and SDevice decks to support subsequent device simulations, though this line of research remains relatively limited. Our work aims to build a complete TCAD SDE/SDevice code generation flow and release an open-source dataset to foster further investigation in this direction. In summary, LLM-based code generation holds great potential to accelerate the EDA workflow across the entire DTCO iteration process.

\subsection{Agent Flow in EDA}
Agents built upon LLMs provide a flexible paradigm for decomposing complex design tasks into modular, cooperative subtasks. Recent studies have applied LLM-based agents to analog design, such as topology exploration of operational amplifiers\cite{2024-dac-analogllm} and hybrid frameworks that combine LLMs with Bayesian optimization for circuit design\cite{2025-iccad-ADO-LLM}. LLM-based agents have further been investigated in digital design, including recent demonstrations on CPU development with successful tape-out results\cite{2024-dac-chatcpu}. In addition, LLM agents have been explored for generating and optimizing circuit layouts \cite{2025-stdlayout-nv, 2025-tcad-LayoutCopilot}, aiming to improve downstream manufacturing quality and design efficiency. Overall, LLM-driven agent flows have shown strong potential to enhance automation and efficiency in EDA.

\section{methodology}
\subsection{Dataset Construction}
To address the absence of publicly available datasets tailored for LLM applications in the design–technology co-optimization (DTCO) domain, particularly in device design, we develop a semi-automated dataset construction flow, as illustrated in Fig. \ref{fig:overall_flow}(a). In the proposed dataset construction flow, TCAD experts first identify a diverse set of representative device architectures. The selection covers both conventional planar transistors and advanced three-dimensional structures such as FinFETs, nanosheets, and nanowires, as well as their array configurations. To enhance material diversity, the dataset also includes devices based on compound semiconductors such as SiGe and InGaAs in addition to traditional silicon. For each device type, experts develop baseline SDE scripts that specify geometry, doping profiles, and contact configurations. Doping distributions are modeled using both uniform concentrations and physically realistic Gaussian profiles derived from ion implantation processes. To ensure simulation robustness and scalability, the meshing strategy is parameterized to adapt to varying device dimensions, with both axis-aligned and offset meshes incorporated for accurate representation of complex structures. Complementary SDevice scripts are further generated to define boundary conditions, physical models, and biasing schemes. These configurations span traditional drift–diffusion models as well as advanced tunneling-based models, enabling simulations to produce a wide range of electrical characteristics, including transfer characteristics ($I_d-V_g$), output characteristics ($I_d-V_d$), and capacitance–voltage ($C-V$) curves. To further scale the corpus, the baseline scripts are systematically expanded through parameter variation in Sentaurus Workbench, resulting in thousands of executable simulation scripts. 

However, these scripts alone are insufficient for direct use in LLM training, as effective SFT requires paired natural-language queries with corresponding code responses, and manually generating such queries at scale is impractical. To transform these scripts into natural language–to–code training data, we design specialized prompting strategies that leverage a commercial LLM to generate diverse user queries consistent with TCAD semantics and design intent, as illustrated in Fig. \ref{fig:dataset_prompt}(a). To support this process, domain experts modularize the baseline codes, annotate them with descriptive comments on structural and parametric details, and package each script together with its metadata into a structured JSON format. In the prompting strategy, the LLM is explicitly designated as an expert in TCAD simulation and prompt engineering, and is provided with precise instructions regarding the information it receives. The prompt description specifies the global context of the target device or study, the type of simulation such as SDE for structure creation or SDevice for $I_d-V_g$, $I_d-V_d$ and $C-V$ analyses, as well as the form of metadata, which may be given as default parameters, symbolic templates, or explicit numerical values. The task is formulated to require the generation of exactly one concise and natural-sounding query that reflects typical engineering practice, while adhering to strict output format constraints to ensure clarity and consistency. In addition, representative examples are included, and controlled variation in phrasing style is encouraged, thereby capturing the diversity with which TCAD engineers naturally articulate their requests. 

This representation enables the LLM to generate high-quality queries that are consistent with the underlying code. Finally, domain experts conduct manual verification to filter and refine the augmented data, ensuring both physical plausibility and the feasibility of subsequent simulations, resulting in the finalized dataset.



\subsection{Domain-Specific LLM Fine-Tuning}
\begin{figure*}[t]
\centering
\includegraphics[width=0.94\linewidth]{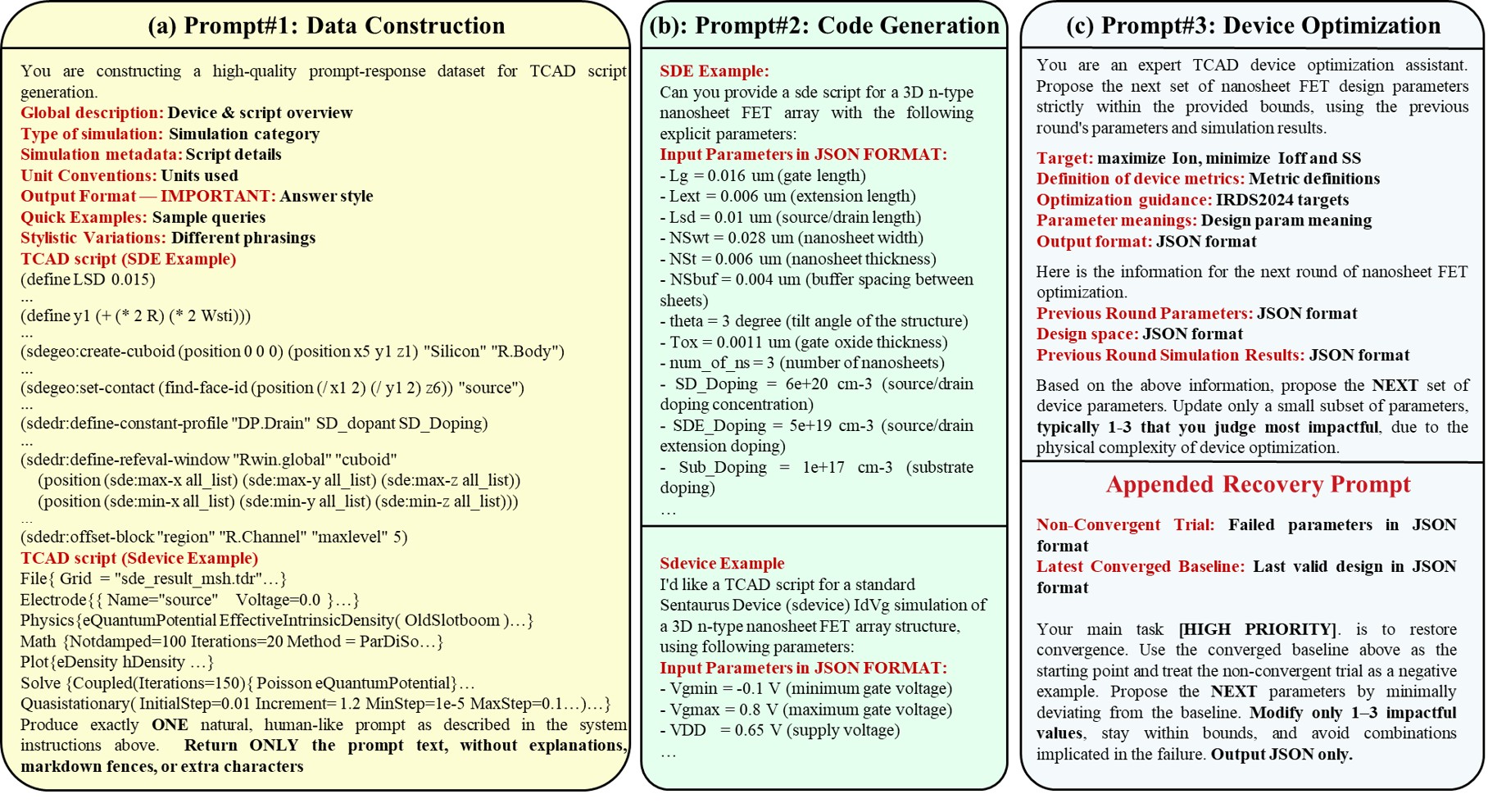}
\caption{Prompt templates and examples used in our framework. (a) Data Construction, (b) Code Generation and (c) Device Optimization.}
\label{fig:dataset_prompt}
\end{figure*}

After constructing the TCAD-specific dataset, we further adapt a general-purpose LLM to the task of code generation through SFT\cite{2021-sft-openai}. In this stage, the pretrained model is exposed to a large number of carefully designed input–output pairs so that it can gradually adjust its parameters toward the characteristics of TCAD-related tasks. The fine-tuning objective is formulated as a next-token prediction problem, where the model is optimized to maximize the likelihood of generating the correct target sequence conditioned on the given input. Concretely, each training example is represented as a pair: the input corresponds to a natural-language query derived from the prompt templates in Fig. \ref{fig:dataset_prompt}(a), while the output is the associated SDE or SDevice script that implements the requested functionality. This design explicitly bridges the gap between high-level textual descriptions and low-level simulation code.

During training, we adopt a causal language modeling objective in which all tokens belonging to the prompt are masked out during loss computation. By doing so, the model does not simply learn to copy the input, but instead concentrates on capturing the correct dependencies required to produce the desired output sequence. This setup encourages the model to internalize domain-specific conventions such as the syntax, parameter formatting, and hierarchical structure of TCAD scripts. As a result, the fine-tuned model becomes better aligned with the practical requirements of semiconductor device simulation tasks, enabling agile and efficient transistor design.

\subsection{Multi-Agent Workflow for Automated Device Optimization}
Building on our domain-specific TCAD code generation LLM, we design \textbf{AgenticTCAD}, a multi-agent workflow that integrates the fine-tuned model with a commercial reasoning LLM, as illustrated in Fig. \ref{fig:overall_flow}(c). Each agent is equipped with a tailored prompting strategy to handle specialized subtasks, thereby enabling a fully automated loop that covers device code generation, TCAD simulation execution, post-processing, and iterative optimization. To validate the proposed framework, we apply \textbf{AgenticTCAD} to the optimization of design parameters for nanosheet field-effect transistors (NS-FETs) at advanced technology nodes. This workflow effectively connects natural language design specifications to the automated optimization of device performance, providing a practical framework for accelerating TCAD-driven device exploration.

In the stage 1, as shown in Fig. \ref{fig:overall_flow}(c), \textbf{AgenticTCAD} employs the fine-tuned code generation LLM to convert user-provided design parameters, expressed in JSON format, into simulation-ready inputs. As illustrated in Fig. \ref{fig:dataset_prompt}(b), these parameters are incorporated into tailored prompting strategies, enabling the TCAD code generation LLM to produce the NS-FET structure description through SDE, which specifies device geometry, doping profiles, contact definition, and meshing. In parallel, the LLM generates the corresponding SDevice script, which defines the physical models, boundary conditions, and solver configurations required for electrical simulations. This stage enables the agent to autonomously design NS-FET architectures with user-specified parameters and generate their corresponding simulation scripts, thereby establishing a consistent and automated foundation for subsequent simulation and optimization.

The workflow in stage 2 executes the TCAD simulation and subsequent post-processing. The SDE script is first executed to construct the NS-FET geometry and corresponding device meshing. The resulting structure is then passed to the SDevice solver, which performs electrical simulations based on the associated physical models, boundary conditions, and solver settings. According to the International Roadmap for Devices and Systems (IRDS)-2024 logic device requirements \cite{IRDS2024}, three key performance metrics are selected, and the post-processing module automatically extracts them from the simulated $I-V$ characteristics. The on-state current ($I_{on}$) is the drive current that determines both the switching speed of logic gates and the drive strength of the device. The off-state current ($I_{off}$) corresponds to the leakage current in the closed state and directly impacts static power consumption and overall energy efficiency. The subthreshold swing ($SS$) quantifies the sharpness of the switching transition and governs the trade-off between performance and power efficiency. These metrics, together with the complete $I-V$ characteristics, are stored in a structured JSON format to enable seamless integration with the subsequent reasoning stage. In addition, the post-processing module also extracs the energy band diagrams in the channel region under different bias conditions. These diagrams include the conduction band ($E_{C}$), the valence band ($E_{V}$), the electron quasi-Fermi level ($E_{Fn}$), and the hole quasi-Fermi level ($E_{Fp}$). Such information provides physical insight into carrier transport and enhances the interpretability of the optimization process.

The results of the first two stages complete the forward process of device simulation, while Stage 3 employs a commercial reasoning LLM as a device optimization agent to realize feedback optimization. The agent receives the simulation results, the corresponding design parameter space, and the previous round of design parameters as references. These inputs are integrated into carefully engineered prompts, as illustrated in Fig. \ref{fig:dataset_prompt}(c), where the performance metrics are supplemented with detailed physical interpretations and the optimization task is explicitly specified. Based on this information, the device optimization agent is instructed to generate updated design parameters for the next round of code generation and simulation. At the same time, it evaluates whether the simulated performance meets the specifications defined by the IRDS-2024 or whether the maximum number of iterations has been reached. Since TCAD inherently involves solving nonlinear semiconductor device equations, convergence issues may arise for certain parameter sets. To address this challenge, \textbf{AgenticTCAD} incorporates a recovery mechanism, as shown in Fig. \ref{fig:dataset_prompt}(c). In cases where non-convergence is encountered, the system provides the agent with the most recent convergent design, the failed configuration, and an additional prompt specifically designed for handling non-convergence. This mechanism enables the reasoning agent to revise the design parameters and ensures that the optimization loop continues without interruption.


\begin{figure}[tp]
\includegraphics[width=1\linewidth]{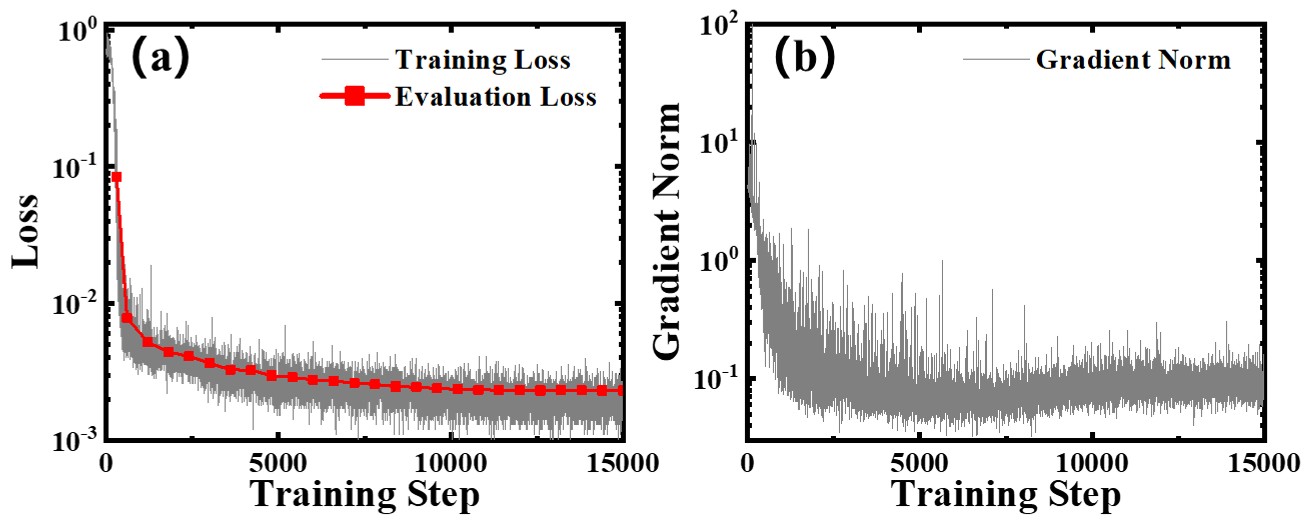}
\caption{SFT process: (a) training and evaluation loss, and (b) gradient norm.}
\label{fig:training_loss}
\end{figure}

\begin{figure}[tp]
\centering
\includegraphics[width=1\linewidth]{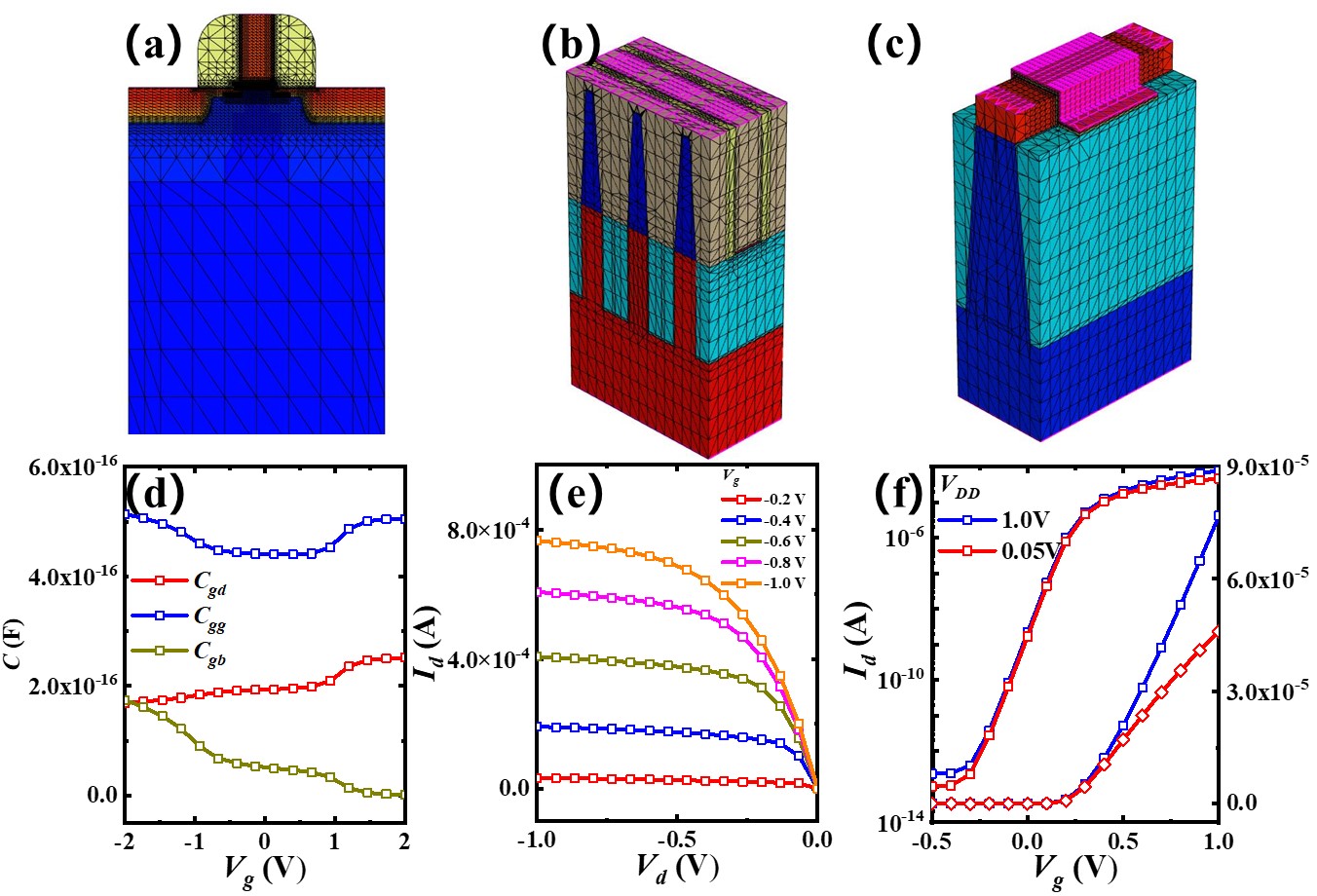}
\caption{Case study results with our TCAD code generation LLM with tailored meshing strategies. (a) 2D n-type MOSFET with Gaussian doping profile and (d) the corresponding $C-V$ simulation. (b) P-type FinFET array and (e) the corresponding $I_d-V_d$ simulation. (c) InGaAs-based n-type FinFET with high-k/metal gate stack and (f) the corresponding $I_d-V_g$ simulation.}
\label{fig:tcad_case}
\end{figure}

\section{Experiments and results}
\subsection{Experimental Setup}
All experiments were conducted on a Linux-based workstation equipped with two Intel Xeon Platinum 8375C CPUs, each operating at 2.9 GHz with 64 cores, and 4 NVIDIA A6000 GPUs with 48 GB of memory. For SFT, we employed the open-source Qwen2.5-14B-Instruct model\cite{Qwen25}, implemented with the Transformers library. The fine-tuning was performed on 30,000 training samples for 5 epochs with a batch size of 8 in bfloat16 precision. The AdamW optimizer was used with momentum parameters $\beta_{1}=0.9$ and $\beta_{2}=0.95$, a weight decay of $1\times10^{-4}$, an initial learning rate of $1\times10^{-5}$, a cosine learning-rate decay schedule, and a warmup ratio of 0.05. The training procedure required approximately 107 hours on four A6000 GPUs. For TCAD execution, we utilized Synopsys Sentaurus W-2024.09-SP1. The code generation model was deployed on a single A6000 GPU, and the device optimization was carried out using DeepSeek V3.1 reasoning version\cite{deepseekv31}.

\subsection{Open TCAD Dataset and Model Evaluation}
Fig. \ref{fig:training_loss} summarizes the SFT process for TCAD code generation. The training loss in Fig. \ref{fig:training_loss}(a) decrease consistently, with periodic evaluations confirming the absence of overfitting. The gradient norm shown in Fig. \ref{fig:training_loss}(b) gradually declines as the parameters are updated, indicating stable convergence and effective acquisition of domain-specific knowledge. Representative case studies are presented in Fig. \ref{fig:tcad_case}, where the fine-tuned model generates both device structures and corresponding simulations. The 2D n-type MOSFET in Fig. \ref{fig:tcad_case}(a) includes a Gaussian doping distribution to emulate ion implantation, and the generated script produces the expected $C-V$ characteristics in Fig. \ref{fig:tcad_case}(d). The p-type FinFET array shown in Fig. \ref{fig:tcad_case}(b) yields valid $I_{d}–V_{d}$ results at different $V_g$ in Fig. \ref{fig:tcad_case}(e). Moreover, The InGaAs-based n-type FinFET with a high-$k$ metal gate stack in Fig. \ref{fig:tcad_case}(c) generates both linear and saturated $I_{d}–V_{g}$ curves in Fig. \ref{fig:tcad_case}(f). These results demonstrate both the diversity of our dataset and the ability of SFT to enable reliable domain-specific TCAD code generation.

Table \ref{tab1:compareSOTA} compares our framework with the state-of-the-art approaches. Existing studies on TCAD code generation remain largely focused on the SDE stage with preliminary explorations, and the community still faces the bottleneck of lacking diverse datasets. In contrast, we provide an expert-crafted TCAD dataset and establish a complete natural language–driven framework that employs a domain-specific TCAD LLM as the code generator to cover device structure and electrical simulation, and further integrates a closed-loop device optimization stage, thereby pushing this research area forward.

\begin{table}[tbp]
\centering
\caption{Comparison of TCAD code generation approaches.}
\begin{tabular}{cccccc}
\toprule
\multirow{2}{*}{\textbf{Functionality}} 
 & \multicolumn{2}{c}{\textbf{SDE Code}} 
 & \multicolumn{2}{c}{\textbf{Sdevice Code}} 
 & \textbf{Device Agent} \\
\cmidrule(lr){2-3} \cmidrule(lr){4-5} \cmidrule(lr){6-6}
 & Structure & Mesh & I-V & C-V & Optimization \\
\midrule
{\cite{2024-nano-tcadllm}} & \checkmark & $\times$ & $\times$ & $\times$ & $\times$ \\
MALTS\cite{2025-lay2sde} & \checkmark & \checkmark & $\times$ & $\times$ & $\times$ \\
ChatTCAD\cite{2025-chattcad} & \checkmark & \checkmark & $\times$ & $\times$ & $\times$ \\
AgenticTCAD & \checkmark & \checkmark & \checkmark & \checkmark & \checkmark \\
\bottomrule
\end{tabular}
\label{tab1:compareSOTA}
\end{table}

\subsection{Multi-Agent Workflow Evaluation}
To evaluate the effectiveness of the proposed multi-agent workflow, we apply \textbf{AgenticTCAD} to the design and optimization of a NS-FET at advanced technology nodes, which represents a key direction for next-generation transistor architectures. Guided by the high-performance specifications defined in IRDS-2024 for the 2 nm technology node with a supply voltage of 0.65 V, the framework aims to adjust device parameters such that the resulting performance meets design requirements.

\begin{figure}[tbp]
\includegraphics[width=1\linewidth]{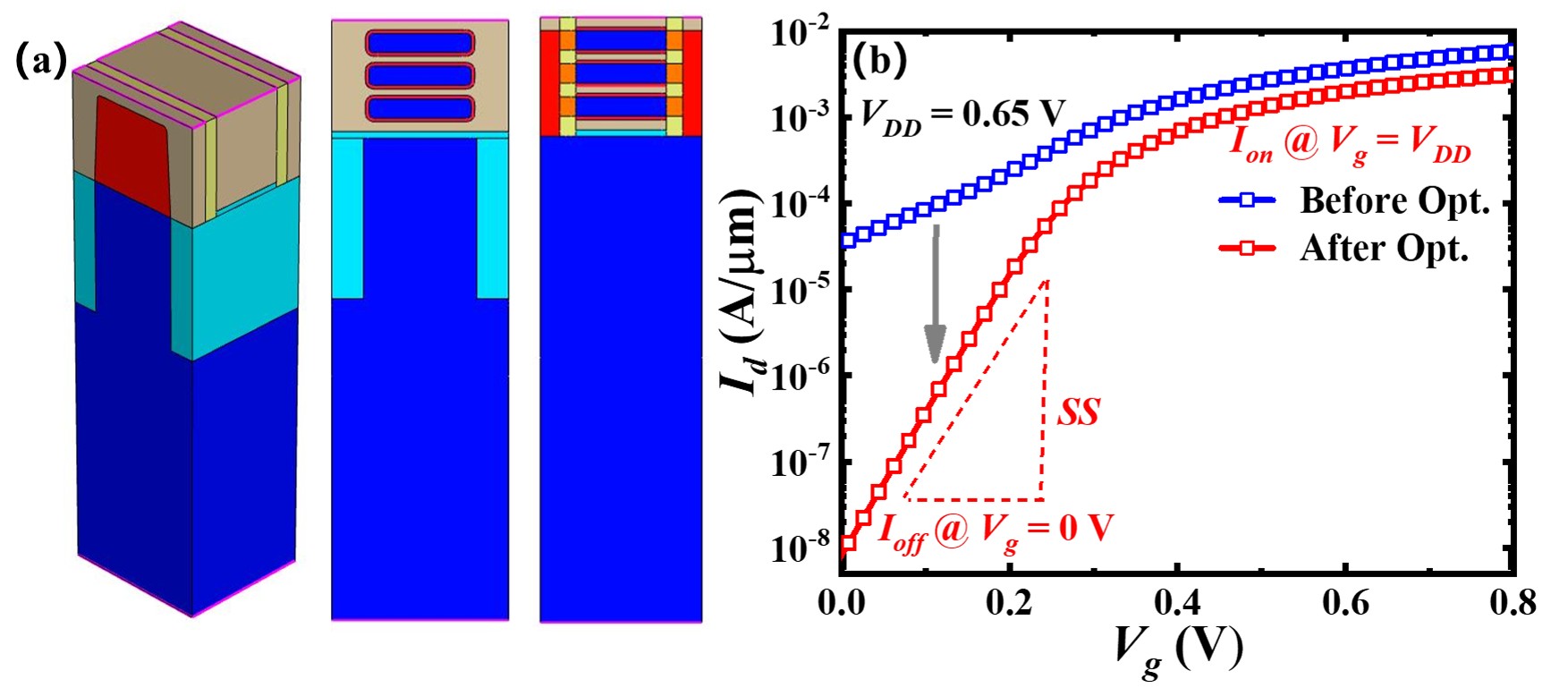}
\caption{TCADAgent for NS-FET design and optimization: (a) device structure and cross-section, (b) $I_d-V_g$ before (blue) and after (red) optimization.}
\label{fig:design_result}
\end{figure}

Fig. \ref{fig:design_result}(a) illustrates the three-dimensional device structure and cross-section automatically generated through the code generation process. Fig. \ref{fig:design_result}(b) shows the corresponding $I_{d}–V_{g}$ characteristics before and after optimization. The blue curve corresponds to an initial design with randomly assigned parameters, while the red curve represents the outcome of our agent-based optimization, which achieves a significant improvement in device performance. Table \ref{tab2:quant_compare} summarizes the quantitative results of \textbf{AgenticTCAD} in comparison with the IRDS-2024 specifications. Although the randomly initialized design parameters yield an $I_{on}$ that already satisfies the requirement, the $SS$ reaches 286.72 mV/dec and the $I_{off}$ exceeds the specification by more than 3 orders of magnitude, indicating unacceptable leakage power and poor switching characteristics. After optimization, \textbf{AgenticTCAD} successfully reduces $I_{off}$ below the target while achieving an $SS$ of 60.38 mV/dec, surpassing the IRDS requirement of 72 mV/dec for 2 nm techonology node and approaching the theoretical limit of silicon-based devices. At the same time, $I_{on}$ is further enhanced to $2.31 \times 10^{-3}$ A/$\mu$m, which is 2.94 times higher than the specified requirement, leading to a significant improvement in the ON–OFF ratio.

\begin{table}[tbp]
\centering
\caption{Optimization results compared with IRDS specifications}
\setlength{\tabcolsep}{2pt}
\begin{tabular}{lcccc}
\toprule
\textbf{Metric} & \textbf{IRDS Spec} & \textbf{Before opt.} & \textbf{After opt.} & \textbf{Meets Spec} \\
\midrule
SS (mV/dec) & 72 ($\downarrow$) & 286.72 & 60.38 & \checkmark \\
$I_\text{off}$ (A/$\mu$m) & $1.0 \times 10^{-8}$ ($\downarrow$) & $3.40 \times 10^{-5}$ & $8.26 \times 10^{-9}$ & \checkmark \\
$I_\text{on}$ (A/$\mu$m) & $7.87 \times 10^{-4}$ ($\uparrow$) & $4.23 \times 10^{-3}$ & $2.31 \times 10^{-3}$ & \checkmark \\
ON-OFF ratio & 4.90 ($\uparrow$) & 2.10 & 5.45 & \checkmark \\
\bottomrule
\end{tabular}
\label{tab2:quant_compare}
\end{table}

\begin{figure}[tbp]
\includegraphics[width=1\linewidth]{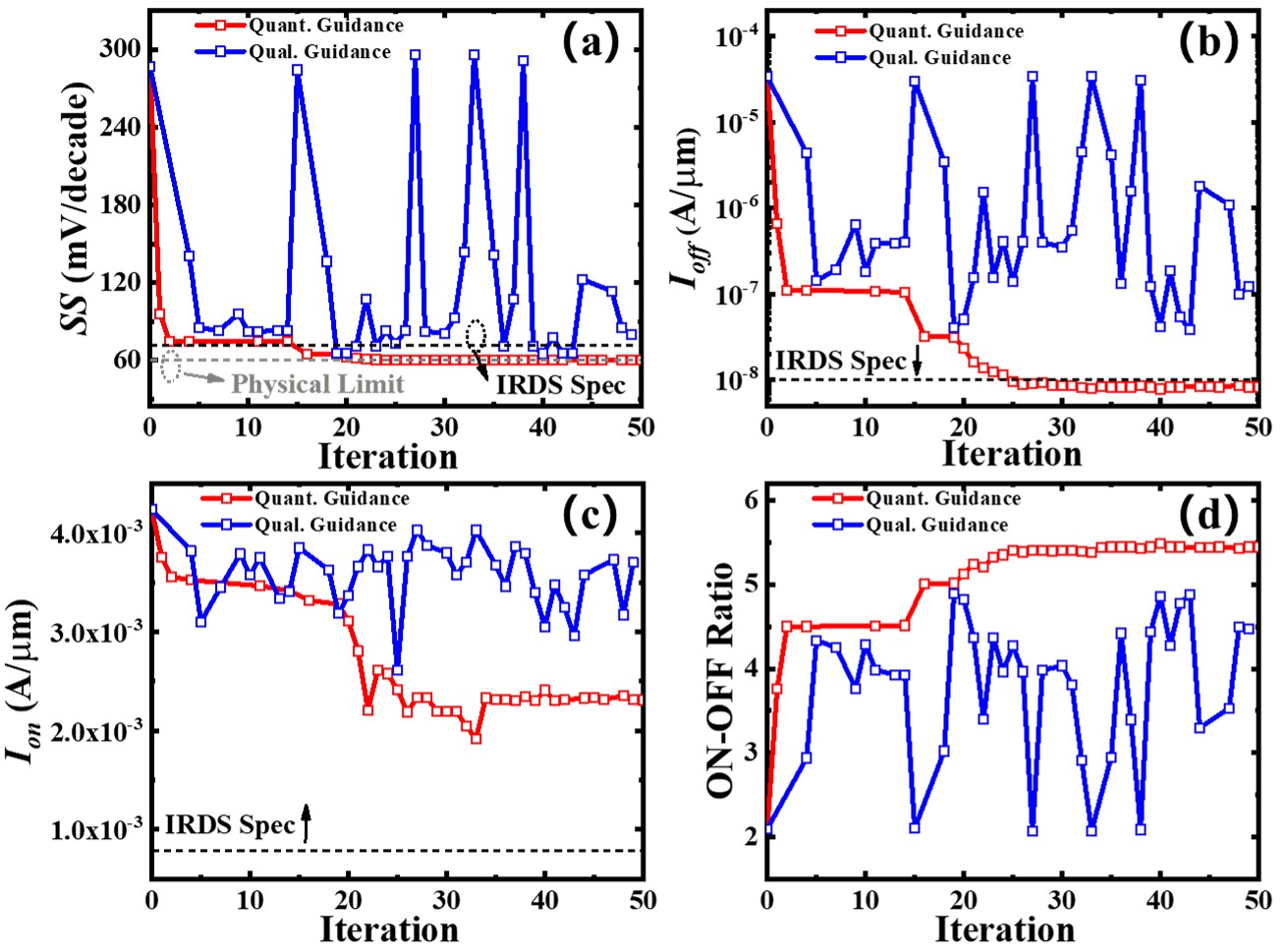}
\caption{Optimization trajectories of device performance under qualitative guidance (blue) and quantitative guidance (red) for (a) $SS$, (b) $I_{off}$, (c) $I_{on}$, and (d) ON-OFF ratio. Missing iteration points correspond to non-convergent design parameters}
\label{fig:device_optimization_process}
\end{figure}

We further examine how different prompting strategies influence the reasoning ability of the device optimization agent. The results are presented in Fig. \ref{fig:device_optimization_process}, which compares optimization trajectories under qualitative and quantitative guidance, shown in blue and red curves respectively. In the case of qualitative guidance, the agent is only informed of the general optimization directions, namely to increase $I_{on}$, to decrease $I_{off}$, and to reduce the $SS$. As illustrated in Fig. \ref{fig:device_optimization_process}(a), the optimization of $SS$ under this setting is unstable. Although the agent is able to approach the target value at certain iterations, the performance often deteriorates in later steps even when the results of previous simulations are provided. A similar trend can be observed in Fig. \ref{fig:device_optimization_process}(b), where $I_{off}$ cannot be effectively improved when the initial value is far from the specification. In this case, the best result obtained through qualitative guidance still exceeds the design requirement by nearly an order of magnitude, which indicates the limitation of providing only directional hints.

When quantitative guidance is adopted, the agent is instead provided with explicit numerical targets relative to the design specifications. This enables stable convergence of the optimization process. As shown in Fig. \ref{fig:device_optimization_process}(a), the $SS$ target is rapidly achieved, and the optimization continues toward values close to the physical limit of silicon-based devices. Since $I_{on}$ and $I_{off}$ are typically in a trade-off relationship, the quantitative setting allows the agent to recognize that $I_{on}$ already exceeds the requirement by a large margin and to strategically sacrifice a small portion of drive current in exchange for significant leakage reduction. This behavior can be clearly observed in Figs. \ref{fig:device_optimization_process}(b) and \ref{fig:device_optimization_process}(c), where $I_{off}$ is successfully optimized even when the initial parameters are highly unfavorable. Furthermore, Fig. \ref{fig:device_optimization_process}(d) shows that the ON–OFF ratio consistently improves during the entire optimization trajectory. This outcome confirms that although some $I_{on}$ is traded off, the overall device performance is substantially enhanced. These results demonstrate that quantitative guidance enables stable and convergent optimization, underscoring the crucial role of prompt design in multi-agent workflows.

\begin{figure}[tbp]
\includegraphics[width=1\linewidth]{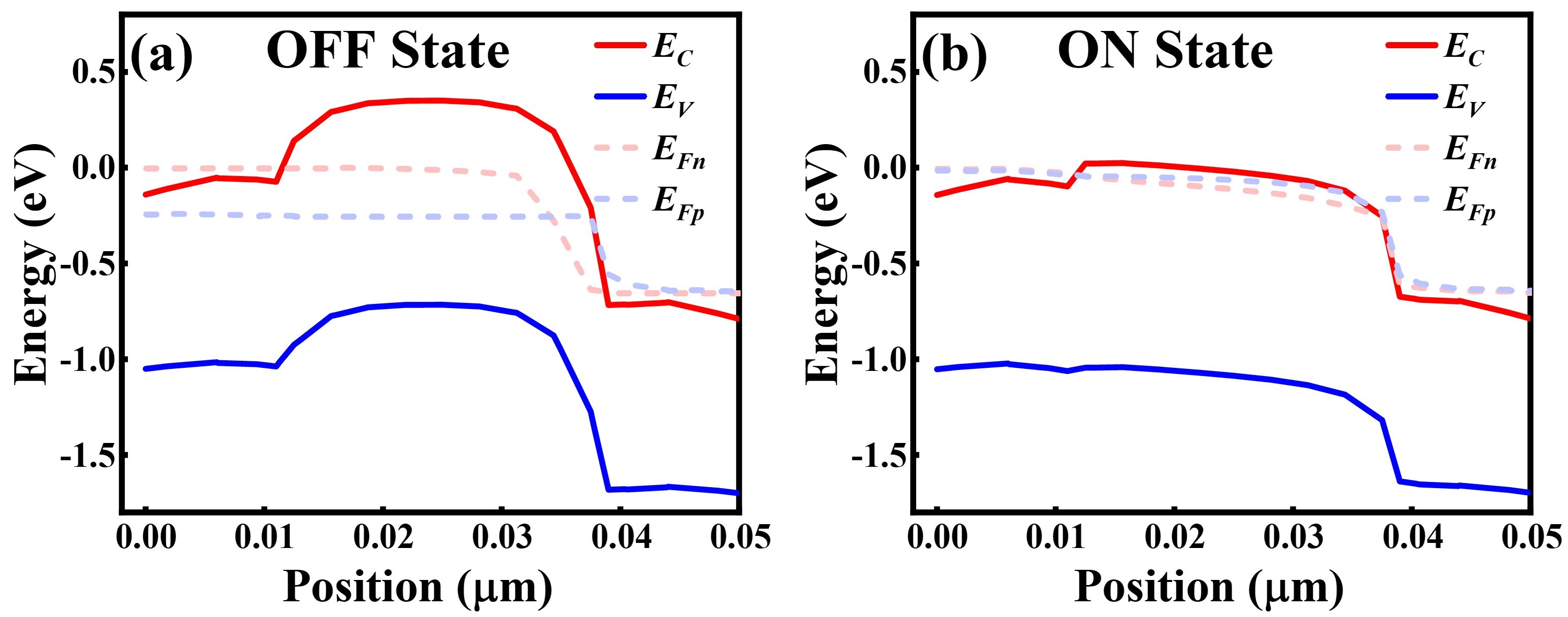}
\caption{Physical insight of NS-FET with energy band diagrams along source-channel-drain in (a) OFF state and (b) ON state.}
\label{fig:banddia}
\end{figure}

In terms of optimization efficiency, Fig. \ref{fig:device_optimization_process} demonstrates that \textbf{AgenticTCAD} requires only 25 iterations to achieve the device performance targets specified in IRDS-2024 for the 2 nm NS-FET, with a total runtime of 4.2 hours. By contrast, a TCAD expert conducted 5,120 simulations using five commercial licenses and required 7.1 days to identify the same design point. These results confirm that \textbf{AgenticTCAD} can substantially accelerate device design and optimization. Furthermore, during the optimization process, the workflow automatically extracts energy band diagrams, as shown in Fig. \ref{fig:banddia}. Under different operating conditions, the diagrams provide direct evidence of carrier transport mechanisms, thereby offering a physically grounded insight for interpreting the optimization trajectory. In this way, \textbf{AgenticTCAD} not only expedites convergence toward target specifications but also enhances the interpretability and transparency of the LLM-based TCAD workflow by explicitly linking quantitative performance improvements to the corresponding underlying device physics.

\section{conclusion}
We have presented \textbf{AgenticTCAD}, a natural language–driven framework for automated device simulation and optimization. Alongside this framework, we constructed an open-source TCAD dataset and fine-tuned a domain-specific language model for TCAD code generation. The effectiveness of \textbf{AgenticTCAD} is demonstrated through the design and optimization of a 2 nm NS-FET, where the framework successfully meets the IRDS-2024 specifications within an end-to-end flow. This contribution marks a significant advance for the TCAD community by establishing a scalable, data-driven pathway to automated device design and optimization and lays the groundwork for future research in interpretable, AI-assisted semiconductor technology development.

\clearpage
\bibliographystyle{IEEEtran}
\bibliography{sample-base}
\end{document}